\documentclass[11pt, a4paper]{article}
\usepackage{moriond,epsfig}
\def\be{\begin{equation}}
\def\ee{\end{equation}}
\def\bea{\begin{eqnarray}}
\def\eea{\end{eqnarray}}

\def\lesssim{\mathrel{\hbox{\rlap{\hbox{\lower4pt\hbox{$\sim$}}}\hbox{$<$}}}}
\def\gtrsim{\mathrel{\hbox{\rlap{\hbox{\lower4pt\hbox{$\sim$}}}\hbox{$>$}}}}

\begin{document}
\vspace*{4cm}
\title{THE TIMESCALE AND MODE OF STAR FORMATION IN CLUSTERS}
%\title{STAR AND STAR CLUSTER FORMATION:\\INSIGHTS FROM THE ORION NEBULA CLUSTER}

% Changed 040623 - All authors go here
\author{JONATHAN C. TAN}
\address{Princeton University Observatory,
Princeton, NJ 08544, USA}

\maketitle\abstracts{ I discuss two questions about the origin of star
  clusters: How long does the process take? What is the mode of
  individual star formation? I argue that observations of Galactic
  star-forming regions, particularly the Orion Nebula Cluster (ONC),
  indicate that cluster formation often takes several Myr, which is {\it
    many} local dynamical timescales. Individual stars and binaries, including massive stars,
  appear to form from the collapse of gas cores.}
% and the stellar mass
%  is mostly determined by the mass of the core. 
%This mode of formation
%  seems to operate even for massive stars.
%As one of the nearest and youngest clusters with a healthy complement
%of massive stars, the Orion Nebula Cluster (ONC) presents a unique
%opportunity for constraining disparate theories of formation.  I
%review what we can learn about the birth of massive stars from the
%BN-KL region, which harbours both an accreting massive protostar and a
%dynamically ejected runaway. I discuss what Orion tells us about
%global models of cluster formation.
\noindent{\small¥ {\it Keywords}: stars: formation --- stars: pre-main sequence --- open clusters and associations}

%\section{Introduction: Two Open Questions}

\section{The Timescale for Star Cluster Formation}

In this section I ask: how long, in terms of
dynamical timescales, does star cluster formation take? In the next, I
consider how gas joins individual stars, forming in a cluster --- is
it from the collapse of pre-existing gas {\it cores} with masses that
help determine the final stellar mass, or from the sweeping-up of gas,
initially unbound to the protostar? I try to derive answers
from an interpretation of observational data and then discuss the
implications for theoretical models.

%first describe some of the theoretical
%background that has motivated different answers to these questions.
%Then I shall try to derive answers from observations, mostly of the Orion
%Nebula Cluster (ONC).

Star clusters are born from the densest gas {\it clumps} in giant
molecular clouds (GMCs). We can measure how long this takes, from the
beginning to end of star formation, in terms of the number of
free-fall timescales, $t_{\rm ff}=(3\pi/32G\rho)^{1/2}$ or the
number of dynamical-crossing timescales $t_{\rm dyn}=R/\sigma =
1.1(G\rho)^{-1/2}$, where $\sigma$ is the 1D velocity dispersion given
by $\sigma^2=\alpha_{\rm vir}GM/(5R)$ with the observed
virial parameter $\alpha_{\rm vir}\sim 1$ (McKee \& Tan 2003). In this
case $t_{\rm ff}=0.50 t_{\rm dyn}$.

The formation timescale is an important overall constraint for
theoretical models. It also determines how much dynamical relaxation
occurs during formation. For $N$ equal mass stars the
relaxation time is $t_{\rm relax}\simeq 0.1 N/({\rm ln} N) t_{\rm
  dyn}$, i.e. about 14 crossing timescales for $N=1000$. Using
numerical experiments, Bonnell \& Davies (1998) found that the mass
segregation time (of clusters with mass-independent initial velocity
dispersions) was similar to the relaxation time.

To evaluate the free-fall and dynamical-crossing timescales requires
a choice of density and/or scale, but in reality density varies with scale.
Gas clumps have approximately
power-law density profiles with $\rho\propto r^{-k_\rho}; \: k_\rho
\simeq 1.5 - 1.8$ (e.g. Mueller et al. 2002).
The stellar distributions of young clusters are also usually centrally
concentrated; e.g. Hillenbrand \& Hartmann (1998) fit King
models to the ONC. Allowing for a 50\% formation efficiency, Elmegreen (2000)
estimated the density in the proto-ONC as $n_{\rm
  H}=1.2\times 10^5$ cm$^{-3}$, so $t_{\rm ff} =
1.25\times 10^5$~yr and $t_{\rm dyn}=2.5\times 10^5$~yr.  This
applies where the present stellar density is
$2\times10^3\:M_\odot\:{\rm pc^{-3}}$, i.e. about
0.3-0.4~pc from the center. Bonnell \& Davies (1998) estimated a
half-mass radius of 0.5~pc, and using 
$\sigma=2.5\:{\rm km\:s^{-1}}$ derived $t_{\rm dyn}\simeq 2\times 10^5$~yr
(note, they define a crossing time $2R/\sigma$, i.e. $4\times
10^5$~yr).

These estimates are based on the present spatial distribution of
matter, even if allowing for a formation efficiency. We expect that
the initial distribution was probably more extended, since
self-gravity and the decay of turbulent support should lead to
contraction, at least in the early stages of formation.  However,
feedback from star formation could maintain turbulence (Norman \& Silk
1980; see below). Also, dispersal of gas leads to the expansion of the
cluster (Hills 1980).

To measure the actual formation time, a common method is to find the
age spread of individual cluster members; for young clusters this is
only practical for lower-mass stars that have a pre-main-sequence
phase once they have finished accreting (high-mass stars reach the
main sequence while still accreting). Observed positions in the
Hertzsprung-Russell (HR) diagram are compared to theoretical models of
evolution (e.g. Palla \& Stahler 1999, hereafter PS99). Modeling
uncertainties include the choice of deuterium abundance; D burning
swells accreting protostars, so raising the ``birthline'' where they
appear on the HR diagram (Stahler 1988).  A relatively high value of
[D/H]$=2.5\times 10^{-5}$ was adopted by PS99. Increasing the
accretion rate also raises the birthline. PS99 used constant rates of
$10^{-5}\:M_\odot\:{\rm yr^{-1}}$, which may be smaller than expected
in the high pressure environments of forming clusters (McKee \& Tan
2003). Finally the birthline is also influenced by the geometry of
accretion (PS99 assumed spherical as opposed to disk accretion), since
this affects the energetics of the gas just before it joins the
stellar surface (Hartmann 2003). Fortunately, as the initial
contraction from the birthline is quite rapid, the birthline position
mostly affects age determinations $\lesssim 1$~Myr old. The ages of
older stars tend to be estimated more robustly. Uncertainties also
become larger at the lowest stellar masses.

The observables for each star are luminosity and surface temperature.
Some stars are likely to be unresolved binaries, which
must be allowed for (e.g. PS99).  The surface temperatures and
luminosities of a substantial fraction of ONC stars have been measured
(Hillenbrand 1997). This sample is biased against very embedded
sources and low-mass stars ($\lesssim 0.4M_\odot$). Patchy extinction
introduces errors. Some systematic errors are evident as the high-mass zero
age main sequence lies above the theoretical expectation by 0.3~dex in
luminosity (or below by 0.05~dex in temperature). If these effects
also operate at lower masses, then ages would be underestimated.

PS99 could estimate ages of 258 stars with masses
$0.4<m_*/M_\odot <6.0$ from Hillenbrand's (1997) sample. The lower
limit of this range was set because of incompleteness at
$L_*\simeq 0.1\:L_\odot$, the predicted luminosity of a
10~Myr old $0.4\:M_\odot$ star.  However, given that the
model uncertainties are largest at these low
masses, and given possible observational systematic uncertainties,
the sample may not be complete at $0.4\:M_\odot$ to ages
of only several Myr. The large numbers of stars at these low masses
increases the potential importance of incompleteness in the lowest
mass bin. When divided into different mass intervals, the lowest mass
bin with mean mass of $\bar{m}_*=0.56\:M_\odot$, does appear to be
different from the next two bins with $\bar{m}_*=1.0,1.8\:M_\odot$.

Assuming a complete sample, PS99 found: 82 stars aged 0-1~Myr, 57 aged
1-2~Myr, 34 aged 2-3~Myr, 17 aged 3-4~Myr, 8 aged 4-5~Myr, 8 aged
5-6~Myr, 8 aged 6-7~Myr, and 6 aged 7-10~Myr. Hartmann (2003) has
argued that the oldest ages ($\sim 10$~Myr) may be due to a problem of
foreground contamination. As described above, birthline uncertainties
mostly affect the youngest ages ($\lesssim 1$~Myr).  We conclude that
a significant fraction of the stars are as old as 3~Myr, and that this
is a lower limit to $t_{\rm form}$ since star formation is still
continuing in the cluster.

An independent method of dating the cluster comes from the
identification of a dynamical ejection event of 4 massive stars (a
binary and two singles) that appear to have originated from the ONC
about 2.5 Myr ago (Hoogerwerf, de Bruijne \& de Zeeuw 2001). The
central value of the time since this ejection event is about 2.3~Myr
in the analysis of Hoogerwerf et al. (2001); however, if the cluster's
distance of about 450~pc is adopted, then the best estimate for the
age is 2.5~Myr. The identification is based upon the fact that the
extrapolation of the motion of the center of mass of the four stars
from the ejection event to the present day, leads to a predicted
position coincident with the ONC (uncertainties are a couple of pc).
These results imply that 2.5~Myr ago the ONC was already a rich
cluster containing at least four stars of spectral type earlier than
O9/B0. Before this the stars had to form and have enough time to find and
eject each other in a close interaction. Thus the estimate of 2.5~Myr
is again a lower limit to $t_{\rm form}$.

Taken together, the above results 
%from pre-main-sequence dating and
%from the time of a dynamical ejection event from the ONC, 
lead us to conclude, with some confidence, that the formation timescale for the cluster is $t_{\rm form}
\geq 3$~Myr. This is $\geq$12 (24) dynamical-crossing (free-fall)
timescales, as estimated at the conditions considered by Elmegreen
(2000) (see above). This age corresponds to a free-fall timescale for
densities of $n_{\rm H}=210\:{\rm cm^{-3}}$, much smaller than the
mean density of the ONC region. The dynamical ejection
event implies that a dense stellar cluster existed 2.5~Myr ago, i.e.
that the densities have not changed appreciably during this time, and
the ONC has taken many ($\gtrsim 10$) dynamical-crossing times to form.

How does this result compare to what is know from other star-forming
regions?  Forbes (1996) did not find evidence for an age spread in
NGC~6531, but the analysis was insensitive to timescales shorter than
at least 3~Myr.  Hodapp \& Deane (1993) analyzed L1641, placing twelve
stars in an HR diagram: ten are spread in age from 0-2~Myr and two are
about 6~Myr old. The conclusions that can be drawn from this cluster
are limited due to the small sample, lack of correction for binarity,
and use of relatively old pre-main-sequence tracks.  Palla \& Stahler
(2000) presented what is so far the most extensive and systematic
analysis of young clusters and associations (in addition to the ONC,
they consider Taurus-Auriga, Lupus, Chamaeleon, $\rho$~Ophiuchi, Upper
Scorpius, IC~348 and NGC~2264).  From these results we again conclude
that $t_{\rm form}\gtrsim 3$~Myr. These systems exhibit a large range
of stellar densities, mostly extending to lower values than the
ONC, so that this timescale does not correspond to as many
dynamical-crossing timescales as in the case of the ONC.
%In some systems, such as Taurus-Auriga, the
%morphology of the stellar distribution is much more filamentary and
%irregular, and this is probably related to the fact that with the
%lower density of this system, $t_{\rm form}\simeq t_{\rm dyn}$.
Siess, Forestini \& Dougados (1997) and Belikov et al. (1998)
concluded that the age spread in the Pleiades was $\sim 30$~Myr around
a mean value of about 100~Myr. If true, then this is the largest
measured formation time of local clusters.

If the formation time is many dynamical timescales, then there should
be virialized and heavily embedded clusters at early stages in their
evolution (most clusters analyzed with the HR diagram method are by
necessity optically revealed and probably nearing the end of star
formation). This population of embedded clusters probably corresponds
to the population of hot molecular clumps observed in sub-mm continuum
and molecular lines (e.g. Mueller et al. 2002; Shirley et al. 2003).
Indeed most of the CS line maps of Shirley et al. (2003) have
morphologies consistent with quasi-spherical virialized distributions.
The clumps have typical masses $M\sim 100-10^4\: M_\odot$, diameters
$\sim 1$~pc and surface densities $\Sigma\sim 1\:{\rm g\:cm^{-2}}$.

Relatively slow, almost quasi-static evolution of the star-forming
clump, contrasts with some theories and models of star formation that
take only one or a few crossing times (e.g. Elmegreen 2000; Bonnell \&
Bate 2002). One theoretical motivation for such fast timescales has
been the rapid decay of turbulence: even in a strongly magnetized
medium, the kinetic energy associated with supersonic turbulence
decays with a half-life of just over a signal crossing timescale
(Stone, Ostriker, \& Gammie 1998). Thus a long formation time requires
that the turbulence observed in star-forming clumps is maintained by
energy input, most likely from protostellar outflows and the overall
contraction of the clump. In fact the energy requirements are quite
modest; the energy dissipation rate of virialized clumps (with
$k_\rho=3/2$ and mean surface density $\Sigma = M/(\pi R^2)$) that
lose half their kinetic energy in one dynamical crossing time, $t_{\rm
  dyn}$ is $21 (M/4000M_\odot)^{5/4} \Sigma^{5/4} \:L_\odot$. Even
very inefficient coupling of protostellar outflows to the gas allows
turbulence to be maintained. The initial kinetic energy of outflows is
expected to be about half of the total energy release associated with
accretion (e.g. Shu et al. 2000), but much of
this is dissipated in shocks as gas is swept-up.  The collimated
nature of the flows also means that a lot of energy escapes from
the star-forming region. Nevertheless, even if only 1\% of
the outflow energy generates turbulence in the
clump, then the cluster can form leisurely in 20 dynamical-crossing
times and still maintain its turbulent support~\footnote{This
  calculation assumes 50\% star formation efficiency, a turbulent
  decay half-life of 1~$t_{\rm dyn}$, a Salpeter initial mass function
  (IMF) from 0.1 to 120~$M_\odot$, protostellar sizes based on the
  model of McKee \& Tan (2003), outflow mechanical luminosity of
  $Gm_*\dot{m}_*/(2r_*)$, and a coupling efficiency of 0.01. The
  formation time that maintains turbulent support is then $t_{\rm
    form} = \eta_{\rm dyn} t_{\rm dyn}$ with $\eta_{\rm
    dyn}=21(M/4000M_\odot)^{-1/2}\Sigma^{-1/2}$.  }. This inefficiency
of star formation is consistent with some numerical simulations of
self-gravitating, unmagnetized gas with driven turbulence
(V\'azquez-Semadeni, Ballesteros-Paredes, \& Klessen 2003).

Note that while the observed age spreads of clusters imply formation
times long compared to the local dynamical timescale of the
clump, these times are about equal to the dynamical timescale of the
larger scale GMCs, in which clumps are embedded. Hartmann,
Ballesteros-Paredes, \& Bergin (2001) used this fact to argue that
GMCs are transient objects, which would be true if GMCs did not
have a long period of quiescence before star formation and if they
were destroyed quickly once the stars had formed. However, the star
formation in most GMCs is spatially localized, so that most of the
mass is quiescent, and it is not clear that most newly formed star
clusters are able to destroy their parental clouds. The observational
study of Leisawitz, Bash \& Thaddeus (1989) found that open star
clusters older than about $\sim 10$~Myr were not associated with
molecular clouds, which is consistent either with post-star-formation
cloud lifetimes shorter than this age or with relative velocities of
star clusters and their parent clouds of about $10\:{\rm km\:s^{-1}}$,
as might arise from photoionization feedback (Williams \& McKee
1997).

One traditional objection to longer formation timescales has been the
idea that once massive stars are formed in a cluster, they would very
rapidly disperse the gas with their ionizing radiation and stellar
winds. Observational evidence suggests that massive stars are not
always the last stars to form in their clusters (e.g. Hoogerwerf et
al. 2001). Tan \& McKee (2001; 2004) found that the turbulent, clumpy,
and self-gravitating nature of gas led to much slower dispersal times,
compared to a uniform medium. A single O star producing $10^{49}$
H-ionizing photons $\rm s^{-1}$ was unable to disperse the gas in a
4000~$M_\odot$ clump with a density comparable to the proto-ONC. If
the gas in the clump was allowed to form stars at a rate such that
50\% of the initial mass would be turned into stars in 15~$t_{\rm
  dyn}$ and the feedback from this star formation was accounted for,
then the gas was dispersed in about 2~Myr (about 10~$t_{\rm dyn}$).
The dynamical ejection of massive stars from clusters, which was not
included in the above calculations, would increase the dispersal
timescale. We have seen that such an ejection occurred in the ONC
2.5~Myr ago. It also appears to be happening at the present epoch with
the ejection of the Becklin-Neugebauer (BN) object and $\Theta^1C$,
the most massive star in the cluster (Plambeck et al. 1995; Tan 2004a).

A 3~Myr formation time for the ONC has implications for how much mass
segregation has occurred, which bears upon the question of whether
massive stars tend to form preferentially in the centers of clusters.
This time corresponds to about 8 diameter crossing times at the half
mass radius of 0.5~pc, which is the unit of time used in the study of
Bonnell \& Davies (1998). If the 30 most massive stars are initially
placed at the half-mass radius, $r_{1/2}$, then after this time the
median location of the 6 most massive stars has migrated in to only
$0.15r_{1/2}$. However, the effects of gradual formation and the
presence of gas need to be accounted for before detailed comparison is
made to the present-day ONC stellar distribution.

\section{The Mode of Star Formation in Star Clusters}

The paradigm of star formation from initially quasi-hydrostatic gas
cores (e.g. Shu, Adams, \& Lizano 1987) has faced the challenges of
competitive accretion and dynamical interactions when applied to star
formation, particularly of high-mass stars, in nascent star clusters
(e.g. Bonnell, Bate, \& Zinnecker 1998; Stahler, Palla, \& Ho 2000).
For example, the smooth-particle-hydrodynamics (SPH) simulations of
Bonnell, Bate and collaborators are characterized by a much more
chaotic evolution in which long-lived cores are not readily apparent.
Bonnell, Vine, \& Bate 2004 showed that the most massive star at the
end of their simulation had gained mass that was initially very widely
distributed.  However, cores (both starless and with embedded
protostars) are observed in real proto-clusters (e.g. Testi \& Sargent
1998; Motte et al. 2001; Li, Goldsmith, \& Menten 2003;
%Walsh et al. 2003;
Beuther \& Schilke 2004), and
even have an IMF that, although uncertain, is similar to that of stars.
%, at least for masses $\lesssim 5\:M_\odot$. 
Part of the resolution of this discrepancy may lie in the methodology
of the numerical simulations: these do not yet include magnetic fields
or feedback from forming stars; the equation of state is isothermal;
the SPH method uses a softening length for each particle so does not
resolve features smaller than this scale, particularly shocks; and
protostars are modeled as sink particles that form when a mass of
$\geq 0.1\:M_\odot$ happens to be self-gravitating, sub-virial and at
a density $n_{\rm H_2}\geq 3.1\times 10^8\:{\rm cm^{-3}}$.  One
example of where the SPH technique may fail is the problem of the rate
of Bondi-Hoyle accretion, which is how most of the stellar mass is
acquired in the above simulations. In this process gas is
gravitationally focused by a passing star so that streamlines collide,
shock and dissipate their energy. It occurs on scales unresolved by
the present SPH techniques. Eulerian grid simulations, including
adaptive mesh refinement of small scale structures, have been used to
simulate the interaction of sink particles with surrounding turbulent
gas: the accretion rate is much smaller than the classical analytic
estimate of accretion from a uniform medium
%, mainly because
%asymmetries in the opposing streamlines lead to reduced dissipation
(Krumholz, McKee, \& Klein 2004).

In spite of recent progress in the observations of high-mass
star-forming regions, the mode of massive star formation remains
controversial. There is as yet no clear-cut example of a massive
protostar for which the accretion disk and core are readily
identified. One difficulty is the uncertainty in the expected
properties of such disks and cores. McKee \& Tan (2003)
described the properties of virialized cores in pressure
equilibrium with their high-pressure proto-cluster surroundings, and
calculated the evolution of the protostars that form from them. Tan
(2004b) compared these models to observations of the Orion Hot Core in
the Kleinmann-Low region of the ONC, concluding that the picture of
massive star formation from the collapse of massive cores was
supported. Here we highlight the main points of this argument.

The total luminosity of the Kleinmann-Low region is about
$10^5\:L_\odot$, and the orientation of the polarization vectors of
$3.8\:{\rm \mu m}$ emission suggest that much of this comes from a
very localized region close to the center of a dense and hot molecular
core, known as the Orion Hot Core (Werner, Dinnerstein, \& Capps
1983). Thermal emission from dust in this structure is also seen at
450~$\rm \mu m$ and it has a radius of $\sim$0.05~pc (Wright et al.
1992), about that expected for an initially 60~$M_\odot$ core.  If
about half of the total luminosity of the region comes from a single
protostar, then it would have a mass of about 20~$M_\odot$ (McKee \&
Tan 2003), so, allowing for accretion inefficiencies due to outflows,
the core is about half way through its collapse.  If the initial ratio
of rotational to gravitational energy is about 2\%, as is typical for
low-mass cores (Goodman et al. 1993), then the expected size of the
accretion disk is about 1000~AU. This corresponds to the size of the
region of SiO (v=0) maser emission seen by Wright et al. (1995). If
the velocity gradient across this structure is interpreted as being
due to Keplerian motion of a disk, then the central mass is about
30~$M_\odot$. The orientation of the disk is perpendicular to the
larger scale molecular outflow that is ejected to the NW and SE of
this region (e.g. Chernin \& Wright 1996). At the center of the SiO
(v=0) emission is a thermal radio source, known as ``I'' (e.g. Menten
\& Reid 1995). It is elongated perpendicular to the disk (Menten \&
Reid, in preparation), i.e. parallel to the outflow. In particular, the
major axis aligns very closely with the direction towards Herbig-Haro
objects to the NW that show blue-shifted velocities of up to 400~$\rm
km\:s^{-1}$ (Taylor et al. 1986). Allowing for flow geometry, this
corresponds to a total velocity of 1000~$\rm km\:s^{-1}$, which
is the expected initial outflow velocity from a $\sim 20-30\:M_\odot$
protostar, i.e. about its escape speed. The thermal radio source is
likely due to ionized gas: the massive protostar is only able to
ionize a small patch of its outflow, which is dense enough to confine
the radiation, except along the rotation axis. These
``outflow-confined'' HII regions have been considered by Tan \& McKee
(2003): the models explain the elongation and radio spectrum of source
``I''. In summary, there are many independent pieces of evidence that
corroborate a model of star formation involving collapse of a massive
core to an accretion disk, which then feeds a massive ($\sim
20\:M_\odot$) protostar, driving a powerful outflow in the process.
No individual piece of evidence is conclusive, but in their totality a
remarkably consistent picture emerges.

The situation is, of course, somewhat more complicated than the above
description. There are a few other protostars and outflows in the
region, but these seem to have quite modest luminosities and their
apparent proximity to the Orion Hot Core may be due to projection
effects. If there is a physically close second protostar, then this
may indicate that the core is collapsing to form a binary, a
relatively minor extension of the basic model.  In addition, the
collapse of the core is probably perturbed by other cluster members;
e.g., the BN object made a close projected passage to source
``I'' about 500~yr ago (Tan 2004a). On smaller scales of several tens
of AU from the center of source ``I'', SiO (v=1) maser emission is
seen with a morphology approximately in the shape of an X, stretched
along the NW-SE axis (Greenhill et al. 2003).  The densities and
temperatures required for excitation are similar to those expected in
the inner part of the outflow (Tan \& McKee 2003), but the velocity
structure has also been interpreted as evidence for a disk aligned
along this NW-SE axis (Greenhill et al. 2003), which would be
incompatible with the above model. The small velocity differences
($\sim$~tens $\rm km\:s^{-1}$) in the maser spots are hard to
understand in the context of an outflow from a massive star that
should have much faster speeds, although the radial velocity range
surveyed so far is only $\sim 100\:{\rm km\:s^{-1}}$.
Further observations of these features would be very useful to help
resolve this issue.

\section*{Acknowledgments}
I thank organizers of this meeting. This work benefited from helpful
discussions with L. Blitz, I. Bonnell, L. Greenhill, M. Krumholz, C.
McKee, P.  Padoan, F. Palla, \& J. Stone. I am supported by a
Princeton Spitzer-Cotsen fellowship \& NASA grant NAG5-10811.

\end{document}